\documentclass[aps, prl, 10 pt, notitlepage, nofootinbib, twocolumn, letterpaper, preprintnumbers, longbibliography, floatfix]{revtex4-1}
\usepackage{epsfig,amsfonts,mathrsfs,amsmath,amssymb,graphicx,color,slashed,bbm}
\usepackage{amsmath,latexsym,amssymb,graphicx,slashed,hyperref,color,enumerate,url,etoolbox}
\usepackage{multirow}

\definecolor{nicered}{rgb}{.7,.1,.1}
\definecolor{nicegreen}{rgb}{.1,.5,.1}
\definecolor{darkblue}{rgb}{0,0,.5}
\hypersetup{colorlinks, citecolor=nicegreen, linkcolor=nicered, urlcolor=darkblue}

\begin{document}

\title{Imperfect Axion Precludes the Domain Wall Problem}

\author{Yue Zhang}
%\email{yzhang@physics.carleton.ca\\[-2ex]}
\affiliation{Department of Physics, Carleton University, Ottawa, Ontario K1S 5B6, Canada}

\date{\today}

\begin{abstract}
The QCD axion needs not be an exact pseudoscalar for solving the strong CP problem. Its imperfectness can play a profound role cosmologically. We propose effective operators, where the Peccei-Quinn field linearly couples to Standard Model particles, provide a dynamical solution to the domain wall problem that prevails in post-inflationary axion models with discrete symmetry. Such interactions generate a thermal potential that drives the axion field to a universal value throughout the universe at high temperatures thus preventing the birth of domain walls when the QCD potential switches on. We discuss generic conditions for this mechanism to work and several concrete examples. Combining with existing electric dipole moment and fifth force constraints, a lower bound on the axion mass is obtained around $10^{-5}$\,eV. Our findings make a strong case for complementary axion searches with both quality preserving and violating interactions.
\end{abstract}

\maketitle

The QCD axion~\cite{Peccei:1977hh, Peccei:1977ur, Weinberg:1977ma, Wilczek:1977pj} provides the leading solution to the strong CP problem in the Standard Model and is subject to a rich program of terrestrial and astrophysical probes~\cite{Sikivie:2020zpn, Irastorza:2018dyq}.
Embedding the spontaneous $U(1)$ Peccei-Quinn symmetry breaking theory in the context of expanding universe brings about both opportunities and problems~\cite{Marsh:2015xka, DiLuzio:2020wdo}. For ultraviolet complete models with more than one flavor of Peccei-Quinn changed quarks, a discrete symmetry often remains after the breaking of Peccei-Quinn symmetry~\cite{Dine:1981rt, Zhitnitsky:1980he}.
If it occurs after inflation, axion domain walls separating different vacuum field values populate the universe~\cite{Sikivie:1982qv}.
Energy density of domain walls redshifts slower than those of radiation and matter. If long-lived they would be disastrous to the otherwise successful standard Big Bang cosmology.

A common solution to the domain problem is to introduce explicit symmetry breaking effect through the scalar potential of the Peccei-Quinn field~\cite{Sikivie:1982qv, Barr:1992qq, Sikivie:2006ni, Barr:2014vva, Caputo:2019wsd}. 
The explicit breaking effect is usually expected to be highly suppressed for the solution to strong CP problem to remain intact.
Taking the operator $\mathcal{L} = e^{i\delta}\phi^n/M_{\rm pl}^{n-4}$ as example, it contributes an axion potential $V\sim (f_a^n/M_{\rm pl}^{n-4}) \cos(a/f_a+\delta)$ after the spontaneous symmetry breaking.
The phase $\delta$ has no reason to be close to $\bar\Theta_\text{\sc qcd}$ from the regular QCD axion potential. Neutron electric dipole moment thus sets a limit $n \gtrsim 200/\ln(\sqrt{2}M_{\rm pl}/f_a)$, e.g., $n\gtrsim12$ for $f_a=10^{11}\,$GeV~\cite{Dine:2022mjw, Banerjee:2022wzk}.
Given the strong Planck-scale suppression, such an explicit breaking term has little cosmological impact at temperatures well above the QCD scale.
After domain walls appear, it can push the walls toward each other which eventually annihilate away. 
Products from the domain walls are littered around the universe. Excessive axion dark matter production and stellar cooling restrict the axion mass within a rather narrow window~\cite{Kawasaki:2014sqa, Ringwald:2015dsf, Beyer:2022ywc, Chang:1998tb}.

In this letter, we aim for a more clean solution to the axion domain wall problem by precluding their existence in the first place. A key observation is that domain walls are not born immediately after the spontaneous Peccei-Quinn symmetry breaking but have to wait until temperature of the universe cools down to the QCD scale. This separation of scales allows one to imagine the following scenario. At temperatures well above the QCD scale, explicit Peccei-Quinn breaking effect generates a transient potential that allows the axion field to settle down at a universal initial value everywhere. At lower temperatures, this potential weakens and eventually gives way to the QCD axion potential. As a result, the whole universe rolls toward today's vacuum (that solves the strong CP problem) starting from the universal initial value set early on. No domain wall is produced in this case even if the low-energy axion potential still possesses an approximate discrete symmetry. For such a mechanism to work, the necessary conditions are: 1) axion potential generated by explicit Peccei-Quinn breaking term is temperature dependent and more important at higher temperatures; 2) the high temperature potential must not have any discrete symmetry; 3) there should be enough time for the settling down of axion field in the first step to complete. Clearly, the $\phi^n$ term mentioned above cannot fulfill this role for more than one reason. An earlier attempt~\cite{Barr:2014vva} resorts to a high-scale confining force to generate the axion potential above the QCD scale. However, additional model building acrobatics is needed for switching off the potential at lower temperatures otherwise the strong CP problem would strike back.

We point out a unique class of explicit Peccei-Quinn breaking interactions that can provide a dynamical, less harmful at low energy and still predictive realization of the above idea,
\begin{equation}
\mathcal{L} = \phi \mathcal{O}_{\rm SM} + {\rm h.c.} \ ,
\end{equation}
where $\mathcal{O}_{\rm SM}$ is a gauge-invariant operator made of Standard Model fields. 
An obvious choice for $\mathcal{O}_{\rm SM}$ would be a term in the Standard Model Lagrangian which is automatically neutral under the Peccei-Quinn symmetry.
At high temperatures the Standard Model particle plasma contributes a potential of the form $\delta V \sim T^4 \cos(a/f_a)$, whereas the zero-temperature counterpart is more suppressed, with $T$ replaced by one of the known particle mass scales. 
Crucially, the high-temperature potential has no discrete symmetry as long as the above interaction is linear in the Peccei-Quinn field $\phi$. 

For concreteness, we present several working examples for realizing the above idea, where $\mathcal{O}_{\rm SM}$ is one of the regular Yukawa interaction operators for fermion mass generation.
First consider the electron Yukawa case,
\begin{equation}\label{eq:2}
\mathcal{L} = \frac{e^{i\delta}}{\Lambda} \frac{\sqrt{2} m_e}{v} \phi \bar L_e H e_R  + {\rm h.c.} \ ,
\end{equation}
where $v=246\,$GeV, $L_e$ is the $SU(2)_L$ lepton doublet, and $H$ is the Higgs doublet. For generality, we include a phase factor $\delta$ that differs from $\bar\Theta_\text{\sc qcd}$ by $\gg10^{-10}$. Hereafter, we focus on the physics after spontaneous Peccei-Quinn symmetry breaking with the radial mode of $\phi$ integrated out, such that $\phi= (f_a/\sqrt2)e^{a/f_a}$.

\begin{figure}[h]
  \begin{center}
    \includegraphics[width=0.4\textwidth]{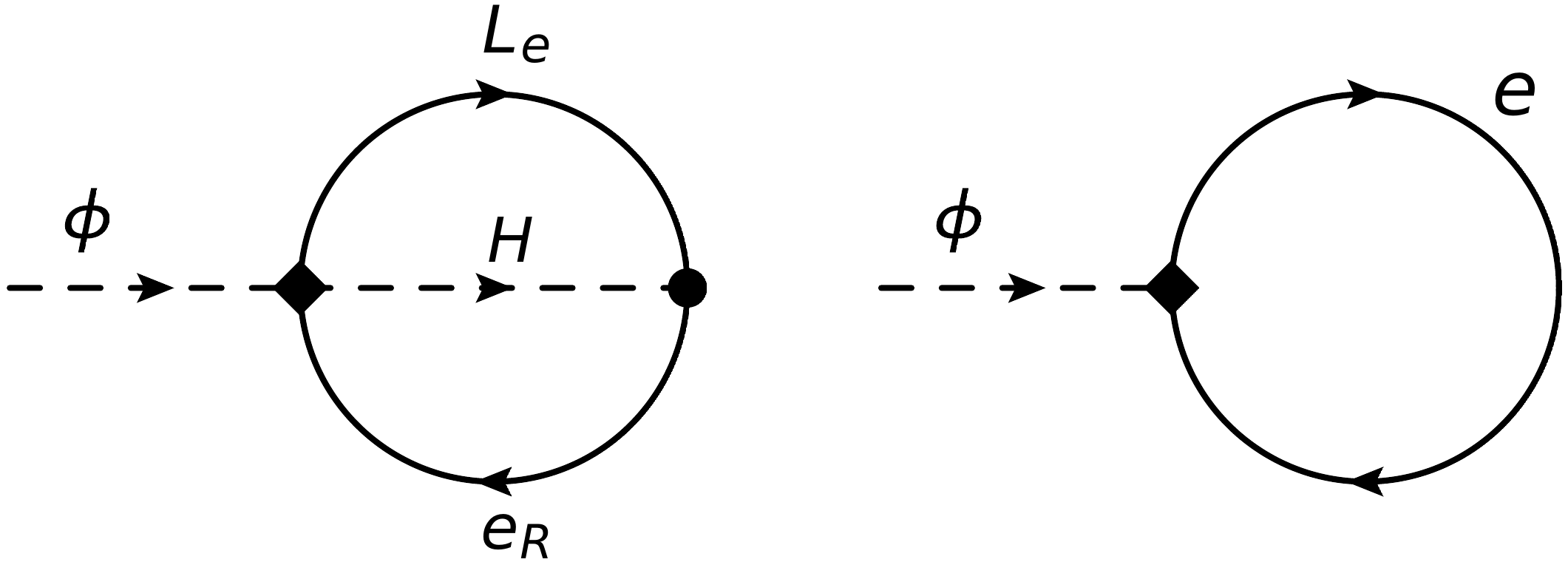}
  \end{center}
  \caption{Feynman diagrams that generate an axion field potential at high temperatures before (left) and after (right) electroweak symmetry breaking. The diamond vertex is given by the operator in Eq.~\eqref{eq:2} or~\eqref{eq:9} whereas the dot vertex is the regular fermion Yukawa coupling. In the right diagram, the diamond contains the Higgs condensate.}
\label{fig:Feynman}
\end{figure}

A thermal potential of axion can be generated by the explicit breaking term Eq.~\eqref{eq:2}, as shown by the diagrams in Figure~\ref{fig:Feynman}. Because the Standard Model particles are in the thermal plasma, the potential can be derived using the imaginary time formalism~\cite{Quiros:1999jp}. At $\Lambda^{-1}$ order,
\begin{equation}\label{eq:3}
V(a, T) = \left\{\begin{array}{cc}
\displaystyle{-\frac{5 m_e^2 f_a T^4}{288\sqrt{2} v^2 \Lambda}} \cos\left( \frac{a}{f_a} + \delta \right), &T\gg v \ ;\vspace{2mm} \\
\displaystyle{-\frac{m_e^2 f_a T^2}{3\sqrt{2}\Lambda} }\cos\left( \frac{a}{f_a} + \delta \right), &\Lambda_{\rm QCD} \ll T < v \ , \\
\end{array}
\right.
\end{equation}
where all loop particles are treated massless.
This serves as the only axion potential at temperatures well above the QCD scale. The axion field's evolution follows the equation
\begin{equation}\label{eq:4}
\ddot a + 3 H \dot a + \partial_a V(a, T) = 0 \ .
\end{equation}
The high-temperature potential drives the axion field value toward the minimum with $a/f_a + \delta =0$, whereas the Hubble parameter exerts damping effect.
We focus on the zero mode evolution in each early patch of the universe that is causally disconnected from others and could start with all possible initial axion field values. Effects due to inhomogeneities are discussed near the end of the paper.
To solve the domain wall problem using the mechanism described above, the axion field must roll sufficiently close to the bottom of the above potential regardless of its initial condition.
The temperature (time) window for this to occur is from the spontaneous Peccei-Quinn symmetry breaking scale to the QCD scale.

\begin{figure*}
  \begin{center}
    \includegraphics[width=1\textwidth]{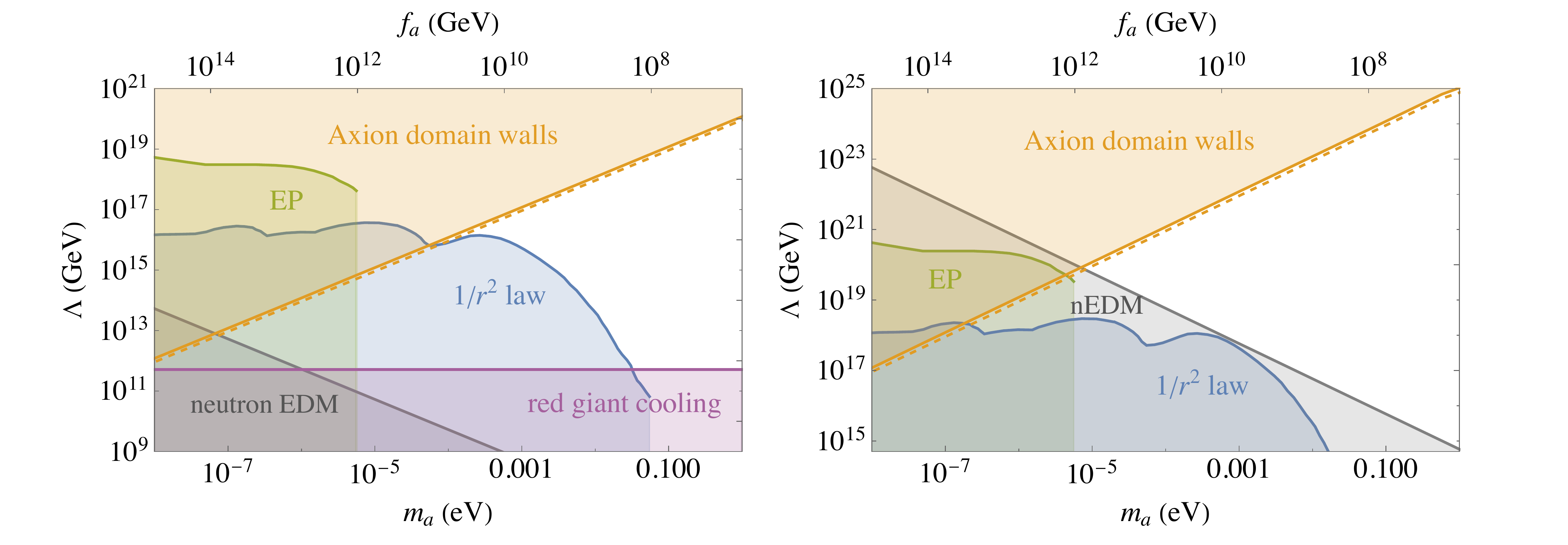}
  \end{center}
  \caption{Parameter space of explicit Peccei-Quinn symmetry breaking effect from the operators Eq.~\eqref{eq:2} (left) and \eqref{eq:9} (right, strange quark case).
  The axion domain wall problem is avoided below the orange shaded region using the mechanism suggested in this work. 
  The solid (dashed) orange curve corresponds to a low energy axion potential with $Z_N$ discrete symmetry with $N=2\,(6)$.
  Lower bounds on the cutoff scale $\Lambda$ come from the tests of equivalence principle (green) and inverse square law of gravity (blue), red giant cooling (purple), and the neutron electric dipole moment measurements (gray) which exclude the shaded regions with various colors.
  The electric dipole moment constraint becomes increasingly stronger for heavier fermions. For the strange quark case, we find the lightest possible axion for solving the axion domain wall problem and staying consistent with existing experimental constraints, with mass $\sim10^{-5}$\,eV.}
\label{fig:electron}
\end{figure*}

We introduce $\zeta = \ln R$ where $R$ is the scale factor of the expanding universe. This allows Eq.~\eqref{eq:4} to be written as
\begin{equation}
\frac{d^2}{d\zeta^2} \frac{a}{f_a} + \frac{d}{d\zeta} \frac{a}{f_a} + \frac{c M_{\rm Pl}^2}{\Lambda f_a} \sin\left( \frac{a}{f_a} + \delta \right) = 0 \ ,
\end{equation}
where $M_{\rm Pl}$ is the Planck scale. The dimensionless parameter $c$ is a constant $1.8\times10^{-16}$ for $T\gg v$, whereas for $\Lambda_{\rm QCD} \ll T < v$, $c$ is temperature dependent and given by $\simeq 2.1\times10^{-10}\,{\rm GeV}^{2} e^{2\zeta}/f_a^2$.
We assume that the Peccei-Quinn symmetry is spontaneously broken at $T=f_a$ and set the corresponding $\zeta$ to zero.
Qualitatively, for the axion field to quickly roll to the minimum of the above potential it requires
\begin{equation}
\frac{c M_{\rm Pl}^2}{\Lambda f_a} \gtrsim1 \ ,
\end{equation} 
leading to an upper bound on $\Lambda$ for given $f_a$.
In practice, we take a number of random initial values for $a/f_a$ ranging between 0 and $2\pi$ and numerically evolve above equation from $\zeta=0$ to $\ln(f_a/\Lambda_\text{\sc QCD})$.
The bound on $\Lambda$ is then derived by requiring for all initial conditions,
\begin{equation}\label{eq:6}
\left|\frac{a}{f_a} + \delta - \bar\Theta_\text{\sc qcd} \right| \lesssim \frac{\pi}{N} \ ,
\end{equation} 
at $\zeta= \ln(f_a/\Lambda_{\rm QCD})$. On the right-hand side, the factor $N$ comes about assuming the low energy axion potential from QCD has a $Z_N$ discrete symmetry ($N>1$).
The value $\pi/N$ represents the typical distance of phase $\delta$ from the closest maximum of the QCD axion potential.
It gives a good estimate except that the high temperature minimum is located in the vicinity of one of the zero temperature maxima.
We disregard such fine-tuned arrangements of parameters because they go against the spirit of naturalness for solving the strong CP problem.
Numerically, our result stays robust against order one variations of $\pi/N$.
Eq.~\eqref{eq:6} guarantees that all early patches of the universe end up in the same vacuum thus avoiding the production of any walls.
The resulting upper bound on $\Lambda$ as a function of $f_a$ is shown by the orange curves in the left panel of Figure~\ref{fig:electron}, for $N=2$ and 6 cases.

The explicit Peccei-Quinn symmetry breaking term Eq.~\eqref{eq:2} can affect a number of low-energy probes. At zero temperature, it shifts the electron mass and in turn sources a Coleman-Weinberg potential for the axion,
\begin{equation}\label{eq:7}
V (a, T=0) = - \frac{|M_e(a)|^4}{64\pi^2} \left( \ln \frac{|M_e(a)|^2}{\mu^2} - \frac{3}{2} \right) \ ,
\end{equation}
which adds to the regular QCD axion potential and can lower the axion quality. Here $M_e(a) = m_e \left[ 1 + {f_a} e^{i(a/f_a + \delta)}/(\sqrt{2} \Lambda) \right]$ and $\mu$ is the renormalization scale.
An order one difference between the $\delta$ and $\bar\Theta_\text{\sc qcd}$ phases will result in a nonzero contribution to the neutron electric dipole moment and requires
$\Lambda > 5.3\times 10^{10}\,{\rm GeV} \left( {10^{-5}\,\rm eV}/{m_a} \right)$~\cite{Zhang:2022ykd}.

In the presence of the explicit Peccei-Quinn breaking, the axion is no longer a pure pseudoscalar boson.
It induces a scalar coupling to electron, allowing the axion to mediate a ``fifth force'' between macroscopic objects. 
Parameters that control the violation of equivalence principle and the inverse square law of gravity are~\cite{Zhang:2022ykd}
\begin{equation}\label{eq:8}
\tilde \alpha = \frac{m_e^2 \sin^2(\delta -\bar\Theta_\text{\sc qcd})}{8\pi G u^2 \Lambda^2}, \quad \alpha = \frac{Z_1 Z_2}{A_1 A_2}\frac{m_e^2 \sin^2(\delta -\bar\Theta_\text{\sc qcd})}{8\pi G u^2 \Lambda^2} \ ,
\end{equation}
where $G$ is the gravitational constant, $u=931.5\,$MeV, $Z_{1,2}$ and $A_{1,2}$ denote the atomic charge and weight of the gravitating objects. 
Existing torsion balance experimental constraints on $\tilde \alpha$ and $\alpha$ are presented in~\cite{Adelberger:2009zz}.
In addition, heavier axions with scalar coupling to electrons are constrained by stellar cooling~\cite{Hardy:2016kme}.
In Figure~\ref{fig:electron}, experimental lower limits on $\Lambda$ are depicted, along with the upper bound for addressing the domain wall problem. 
In this case, fifth force constraint dominates and we obtain a lower bound on the axion mass
$m_a \gtrsim 10^{-4}\, {\rm eV}$.

Next, we move on to the coupling of Peccei-Quinn field directly to quarks as the source of explicit breaking, 
\begin{equation}\label{eq:9}
\mathcal{L} = \frac{e^{i\delta}}{\Lambda} \frac{\sqrt{2} m_q}{v} \phi \bar Q H q_R  + {\rm h.c.} \ ,
\end{equation}
For up-type quark, $H$ is replaced by $i\sigma_2 H^*$.
At high temperatures above the QCD scale, thermal loops similar to Figure~\ref{fig:Feynman} can generate an axion potential that allows it to roll toward the universal value $a/f_a+\delta=0$ throughout the universe. It is useful to note that parametrically the thermal potential is enhanced compared to Eq.~\eqref{eq:3} by a factor of $(m_q/m_e)^2 N_c$, where $N_c=3$ is the color factor.
The larger coupling makes it relatively easier for the axion to settle down to the above minimum to avoid domain walls.

For the low-energy axion potential, naively one may expected it to be generated at tree level once the quark condensate turns on. However, note the tree-level CP-violating effect of \eqref{eq:9} is to modify the phase of quark masses, which are part of $\bar \Theta_\text{\sc QCD}$. After minimizing the axion potential, they make no net contribution to the neutron electric dipole moment~\cite{Zhang:2022ykd}. Instead, the leading axion potential is again of the Coleman-Weinberg form, i.e., Eq.~\eqref{eq:7} but with the mass of electron replaced by the corresponding quark mass along with a color factor.
It is worth noting that such a potential is proportional to fourth power of the fermion mass. The high power of mass dependence implies that the electric dipole moment constraint will stand out further for heavier quarks. 

Through the $\phi$-quark operator, axion also mediates a fifth force by coupling to the nuclei.
In this case, the parameters $\tilde a$ and $a$ are given by
\begin{equation}
\tilde \alpha = \alpha \simeq \frac{f_q^2 m_N^2 \sin^2(\delta -\bar\Theta_\text{\sc qcd})}{8\pi G u^2 \Lambda^2} \ ,
\end{equation}
where $m_N \simeq 938\,$MeV is the nucleon mass, $N$ represents proton or neutron, and $f_q$ is defined through the hadronic matrix element,
$\langle N |m_q \bar q q| N \rangle = f_q m_N \bar u_N u_N$.
We work in the isospin conserving limit and use values of $f_q$ given in~\cite{Cline:2013gha}.
Compared to Eq.~\eqref{eq:8}, $\tilde a$ and $a$ are enhanced by the square of nucleon to electron mass ratio, but as a constraint on $\Lambda$ the mass effect is linear.

Figure~\ref{fig:electron} (right panel) shows the results of Eq.~\eqref{eq:9} where the Peccei-Quinn field couples to a the strange quark Yukawa term.
The color schemes are the same as before, with lower limit on $\Lambda$ set by various experimental results and upper bound for solving the axion domain wall problem. 
Stellar cooling only constrains $\Lambda$ up to $10^{10}$\,eV in these cases~\cite{Hardy:2016kme} and lies well below the range of $\Lambda$ shown in the figure.
Compared to the electron case (left panel), we find that the viable window for $\Lambda$ is pushed to higher values because the axion coupling is enhanced by the larger fermion Yukawa coupling. 
The neutron electric dipole moment constraint becomes significantly more important than equivalence principle and inverse square law tests of gravity, due to the $m_s^4$ versus $m_s^2$ dependences discussed above.
In the strange quark coupling case, we derive the widest axion mass window with a lower bound
\begin{equation}\label{eq:13}
m_a \gtrsim 10^{-5}\, {\rm eV} \ .
\end{equation}

We explore the couplings of Peccei-Quinn field to other quark and lepton flavors as well, by turning on one coupling in each case.
We find that fifth force constraints are more important for light fermions ($u,d, e$) whereas EDM dominates for heavier fermions. The strange quark case is roughly where the transition occurs.
We explore other dimension-5 operators where the Peccei-Quinn field couples to the Higgs and gauge boson fields.
We also consider $\mu e^{i\delta} \phi H^\dagger H$ which is a renormalizable operator.
For the axion solution to the strong CP problem to work, the $\mu$ parameter needs to be very small. To prevent domain walls $\mu$ is also bounded from below. Combining the two requirements, 
we identify a window for $\mu$ as a function of axion mass. Similar to $\Lambda$ for dimension-5 operators, the window closes at a lower bound on the axion mass.
All the results are listed in table~\ref{tab:summarizing}. 
More calculation details are provided in the supplemental material.
Among all operators considered, the lowest possible axion mass is found for Peccei-Quinn field coupling to the strange or muon Yukawa operator, or the gluonic operator.

\begin{table}[h]
  \centering
  %\vspace{0.5cm}
  \begin{tabular}{|c|c|c|c|c|c|c|}
  \hline
 \multirow{2}{2em}{$\mathcal{O}_{\rm SM}$} &  \multicolumn{6}{c|}{Yukawa} \\
  & $u$& $d$& $s$& $c$& $b$& $t$ \\
  \hline
  $m_a^\text{min}$ (eV) & $10^{-4.3}$  & $10^{-4.4}$ & $10^{-5.2}$ & $10^{-4.0}$ & $10^{-3.5}$ & $10^{-2.0}$  \\
  $\Lambda$ (GeV) & $10^{17.5}$ & $10^{18.1}$ & $10^{20.0}$ & $10^{23.4}$ & $10^{25.0}$ & $10^{29.9}$\\
  \hline
    \hline
  \multirow{2}{2em}{$\mathcal{O}_{\rm SM}$} &  \multicolumn{3}{c}{Yukawa} & \multicolumn{3}{| c |}{other operators}  \\
& $e$& $\mu$& $\tau$& $G_{\mu\nu}G^{\mu\nu}$& $(H^\dagger H)^2$ & \multicolumn{1}{||c|}{$H^\dagger H$} \\
\hline
  $m_a^\text{min}$ (eV) & $10^{-4.3}$ & $10^{-4.9}$ & $10^{-3.7}$ & $10^{-5.3}$ & $10^{-2.8}$ & \multicolumn{1}{||c|}{$10^{-2.8}$} \\
  $\Lambda\, (\mu)$ (GeV) & $10^{15.6}$ & $10^{19.9}$ & $10^{23.6}$ & $10^{25.3}$ & $10^{33.1}$ & \multicolumn{1}{||c|}{$10^{-28.6}$}\\ %$10^{-28.5}\,$
  \hline
  \end{tabular}
  \caption{The lowest possible axion mass and corresponding cutoff scale $\Lambda$ from precluding axion domain walls using the mechanism suggested in this work.
  Results are presented for effective operators where the Peccei-Quinn field linearly couples to various Standard Model gauge invariant operators $\mathcal{O}_{\rm SM}$.
  In the last column, the $\phi H^\dagger H$ operator has dimension 3 and $\Lambda$ is replaced by a dimensionful coupling $\mu$.}\vspace{-0.5cm}
 \label{tab:summarizing}
\end{table}
%A linear coupling of the Peccei-Quinn field to Standard Model gauge invariant operator $\mathcal{O}_{\rm SM}$ can forbid the production of domain walls while preserving the solution to the strong CP problem. 

By precluding the birth of axion domain walls, explicit Peccei-Quinn breaking effects introduced through the directly coupling with Standard Model particles (Eq.~\eqref{eq:9} and \eqref{eq:2}) can open up a much wider window of axion mass compared to the Peccei-Quinn field self-interaction. In the latter case, domain walls still occur and viable axion masses are constrained within a very narrow range around tens of meV by the dark matter relic abundance and supernova cooling~\cite{Kawasaki:2014sqa, Ringwald:2015dsf, Beyer:2022ywc}.

The viable axion mass range found above is consistent with the parameter space where axion comprises all cold dark matter in the universe via the misalignment mechanism~\cite{Preskill:1982cy, Abbott:1982af, Dine:1982ah}.
The initial axion field misalignment value is now set as a by-product of the suggested domain wall problem solution, $a/f_a = \delta - \bar\Theta_\text{\sc qcd}$.
Moreover, the lower part of the axion mass range can be probed by the major axion haloscope and helioscope experiments~\cite{ADMX:2018gho, ADMX:2019uok, Rybka:2014xca, Brubaker:2016ktl, HAYSTAC:2018rwy, Alesini:2019ajt, MADMAX:2019pub, Lee:2020cfj, IAXO:2019mpb}.
In complementarity, the parameter space along the $\Lambda$ direction governs the axion quality and will continue to be tested by the future electric dipole moment~\cite{Abel:2020pzs, TUCAN:2018vmr, nEDM:2019qgk}, fifth force experiments~\cite{Adelberger:2009zz}, and monopole-dipole interactions~\cite{ARIADNE:2017tdd, Crescini:2016lwj}.

To summarize, whether the QCD axion is an exact pseudoscalar has been asked as a phenomenological question in various contexts~\cite{Moody:1984ba, Kamionkowski:1992mf, Holman:1992us, Raffelt:2012sp, Mantry:2014zsa, Bertolini:2020hjc, Okawa:2021fto, Dekens:2022gha}. 
This work explores the cosmological significance of a new class of explicit PQ-symmetry breaking effective operators of the form
$\phi \mathcal{O}_{\rm SM}$ which extends previous solutions to the axion domain wall problem. 
Such an operator generates a thermal potential for the axion at temperatures well above the QCD scale, thus allowing the axion to roll and settle down at a universal field value in all patches of the early universe. After the universe cools further, the QCD axion potential takes over but no domain walls is created even if there remains a discrete symmetry, thanks to the above early focusing effect. The proposed mechanism makes prediction in the axion quality and can be tested by a number of ways, including the neutron electric dipole moment and fifth force experiments.
Existing limits set a lower bound on axion mass around $10^{-5}$\,eV for this mechanism to work. Our solution to domain problem opens up a broad window for cosmological viable axion models and strongly motivates a complementary search of both the axion quality preserving and violating interactions with known particles.
The above mass bound does not apply if the axion model is free from discrete symmetries~\cite{Georgi:1982ph, Lazarides:1982tw, Barr:1982uj, Alonso-Alvarez:2023wig}, or the Peccei-Quinn symmetry is already broken before or during inflation~\cite{Seckel:1985tj, Linde:1985yf, Turner:1990uz, Dvali:1995cc, Redi:2022llj}.

We end by commenting on possible topological defects at very high temperatures, around the spontaneous Peccei-Quinn symmetry breaking scale.
Axion string network could arise because the axion field initial values are random among various early patches of the universe. 
The explicit Peccei-Quinn breaking effect discussed in this work cannot only generate the high-temperature potential for setting a common axion field value
but also render the axion strings unstable. Because the early potential features no discrete global symmetry, any domain walls if ever formed along with the strings are unzipped and destroyed right away~\cite{Vilenkin:1982ks}. All these would happen at temperatures shortly after the spontaneous symmetry breaking. 
The universe afterwards is filled with a common axion field value prior to the QCD-scale temperatures, 
i.e., the possible existence of these topological effects at early times does not affect the success of the suggested mechanism.
Their potential roles in the production of axion relic density and the extra clustering of axion dark matter in structure formation are interesting open questions but beyond the scope of this work.

\begin{acknowledgements}
I acknowledge useful discussions with Gonzalo Alonso-\'Alvarez, Luca Di Luzio, and Miha Nemev\v{s}ek at various stages of this work. 
This work is supported by the Arthur B. McDonald Canadian Astroparticle Physics Research Institute and the Natural Sciences and Engineering Research Council of Canada.
\end{acknowledgements}

\bibliography{References}
\end{document}